\documentclass[a4paper]{jpconf}
\usepackage{graphicx}
\usepackage{amsmath}
\usepackage{amssymb,epsfig}

\newcommand{\bea}{\begin{eqnarray}}
\newcommand{\eea}{\end{eqnarray}}
\newcommand{\be}{\begin{equation}}
\newcommand{\ee}{\end{equation}}
\newcommand{\resetequ}{\setcounter{equation}{0}}

\newcommand{\de}{\delta}
\newcommand{\qed}{{\hfill $\Box$}}

\begin{document}
\title{Exorcizing the Landau Ghost in Non Commutative Quantum Field Theory}
\author{Razvan Gurau}

\address{Laboratoire de Physique Th\'eorique, CNRS UMR 8627,
 Universit\'e Paris-Sud XI, 91405 Orsay}

\ead{Razvan.Gurau@th.u-psud.fr}

\begin{abstract}
We show that the simplest non commutative renormalizable field theory, the $\phi^4$ 
model on four dimensional Moyal space with harmonic potential 
is asymptotically safe to all orders in perturbation theory 
\end{abstract}

\section{Introduction} 

This comunication is based on the paper \cite{beta}

Non commutative (NC) quantum field theory (QFT) may be important for physics 
beyond the standard model and for understanding the quantum
Hall effect \cite{DN}. It also occurs naturally as an effective regime of string theory \cite{CDS,SW}.
It led Connes and Chamesddine \cite{Connes} to reformulate the standard model in terms of a spectral triple on a simple non commutative geometry.

The simplest NC field theory is the $\phi_4^4$ 
model on the Moyal space. Its perturbative renormalizability 
at all orders has been proved by
Grosse, Wulkenhaar and followers \cite{GW1,GW2,RVW,GMRV}. 
Grosse and Wulkenhaar solved the difficult problem of ultraviolet/infrared
mixing by introducing a new harmonic potential term 
inspired by the Langmann-Szabo (LS)
duality \cite{LS} between positions and momenta. 

Many on the techniques of commutative field theory have been generalized to include this model. The parametric representation of this model has been established in \cite{GR}. Dimensional regularization has been performen in \cite{dimreg} and the Complete Mellin representation has been introduced in \cite{melin}. 
The Hopf algebra associated structure was introduced in \cite{ConKrTV}.
Models with more general propagators have been analysed in \cite{propaga}. There parametric representation has been introduced in \cite{param2}
It is now tempting to conjecture that commutative renormalizable theories in general have NC renormalizable
extensions to Moyal spaces which may imply some new parameters. 

Once perturbative renormalization is understood, the next problem is to
compute the renormalization group (RG) flow.
It is well known that the ordinary commutative $\phi_4^4$
model is not asymptotically free in the ultraviolet regime.

It is easy enough to understand this phenomenon in the commutative theory. The coupling is given by the amputed one particle irreducible four point function $\Gamma^4$. To each vertex there correspond two propagators, thus the effective coupling is given by
\bea
\lambda_{i-1}=\frac{\Gamma^4_i}{Z^2_i} \; ,
\eea
with $Z$ the wave function renormalization.
For a theory with UV cutoff $\Lambda$, one can follow the evolution of the effective coupling with the scale by means of the function $\beta(\lambda)$ defined as
\bea
\beta(\lambda)=\frac{d\lambda^{nu}}{d\ln\Lambda}\Big{|}_{\lambda^{ren}=ct.} \; .
\eea
Alternatively, in multiscale analysis one uses the definition
\bea
\beta(\lambda)_i=\lambda_{i-1}-\lambda_i \; .
\eea

For the commutative $\Phi^4_4$ theory, the effective coupling varies with the scale. This phenomenon is easily enough understood at the first order in perturbation theory. For the $\Gamma^4$ function we have a non trivial contribution coming from the bubble graph. On the other hand the tadpole, being local gives only a mass counterterms. and consequently $Z=1$ at one loop.

A detailed study shows that if one wants a non zero renormalized coupling constant one needs to start with a large bare coupling constant. Actually the bare coupling becomes zero for some finite UV cutoff. Conversely, for all finite bare couplings the IR theory is trivial (i.e free). This is a serious problem in commutative field theory and has was baptized the "Landau ghost". The infinite quantities subtracted by renormalization are picked up again in this new divergence.

This problem almost killed the QFT. Being an universal phenomenon, exhibited by many field theories, including electrodynamics, it almost led to abandoning QFT as a reasonable description of fundamental intercations. Fortunately QFT was saved by the discovery of ultraviolet asymptotic freedom in non-Abelian gauge theory \cite{thooft}. But in some sense even the asymptotic freedom is not entirely satisfying. It is more like the ghost turned upsind down: it is the UV theory that becomes trivial.

 It is true that such a theory makes much more sense than a theory with ghost, and this allows  the 
introduction of the standard model for elementary pareticles. Nevertheless IR phenomenon (corresponding to a large coupling non perturbative regime) like quark confinement can not be easily understood. Morover the flows of the three couplings in the standard model do not  converge to a unified constant. In order to achieve the convergence of flows one could for instance introduce supersymmetry, but this is problematic as no detection of any super symmetric partner has ever been made.

This same phenomenon blocks the construction of the commutative  $\Phi^4_4$ model.  One could argue that constructive field theory is more an academic question than a true physical problem, but with out a constructive argument perturbative computations have no meaning. For instance the perturbation series could be sensless starting with the first order! 

It was noted in  \cite{GWbeta} that the non commutative $\phi_4^4$ model does not exhibit any Landau ghost 
at one loop. It is not asymptotically free either. 
For any renormalized harmonic potential parameter $\Omega_{ren} >0$, 
the running $\Omega$ tends to the special LS dual point $\Omega_{bare} =1$ in the ultraviolet. As a result
the RG flow of the coupling constant is simply bounded \footnote{The Landau ghost can be recovered
in the limit $\Omega_{ren}\to 0$.}. This result was extended up to three loops in \cite{DR}.

This is due to the fact that in NCQFT the tadpole is no longer local! We have a non trivial wave function renormalization starting with the first order! Moreover it exactely compensates the bubble contribution in the beta function! If one generalizes such an argument to all orders the theory would be finite but not trivial all along its RG flow! 

In this paper we evaluate the flow at the special LS dual point $\Omega =1$, and prove that 
the beta function vanishes at all orders using a kind of Ward identity.
We think the Ward identities discovered here might be important for the
future study of more singular models such as Chern-Simons or Yang Mills theories.

\section{Notations and Main Result}
\resetequ

We adopt simpler notations than those of \cite{GWbeta,DR}, and normalize so that $\theta =1$,
hence have no factor of $\pi$ or $\theta$.

The  bare propagator in the matrix base at $\Omega=1$ is
\be \label{propafixed}
C_{m n;k l} = C_{m n} \delta_{m l}\delta_{n k} \ ; \ 
C_{m n}= \frac{1}{A+m+n}\  ,
\ee
where $A= 2+ \mu^2 /4$, $m,n\in \mathbb{N}^2$ ($\mu$ being the mass)
and we used the notations
\be
\de_{ml} = \de_{m_1l_1} \de_{m_2l_2}\ , \qquad m+n = m_1 + m_2 + n_1 + n_2 \ .
\ee

There are two version of this theory, the real and complex one. We focus on the complex case, the result
for the real case follows easily \cite{DR}.

The generating functional is:
\bea
&&Z(\eta,\bar{\eta})=\int d\phi d\bar{\phi}~e^{-S(\bar{\phi},\phi)+F(\bar{\eta},\eta,;\bar{\phi},\phi)}\nonumber\\
&&F(\bar{\eta},\eta;\bar{\phi},\phi)=  \bar{\phi}\eta+\bar{\eta}\phi \nonumber\\
&&S(\bar{\phi},\phi)=\bar{\phi}X\phi+\phi X\bar\phi+A\bar{\phi}\phi+
\frac{\lambda}{2}\phi\bar{\phi}\phi\bar{\phi}
\eea
where traces are implicit and the matrix $X_{m n}$ stands for $m\delta_{m n}$. $S$ is the action and $F$ the external sources. 

As before, denote $\Gamma^4(a,b,c,d)$ the amputated one particle irreducible four point function with external indices set to $a,b,c,d$, and $\Sigma(a,b)$ the amputated one particle irreducible two point function 
with external indices set to $a,b$ (also called the self-energy). The wave function renormalization is $Z=1-\partial \Sigma(0,0)$ where $\partial\Sigma(0,0)\equiv\partial_L \Sigma = \partial_R \Sigma = \Sigma (1,0) - \Sigma (0,0)$ is the derivative of the self-energy with respect to one of the two indices $a$ or $b$ \cite{DR}.
Our main result is:

\medskip
\noindent{\bf Theorem}
\medskip
The equation:
\bea\label{beautiful}
\Gamma^{4}(0,0,0,0)=\lambda (1-\partial\Sigma(0,0))^2
\eea
holds up to irrelevant terms \footnote{Irrelevant terms include in particular all non-planar or planar with more than one broken face contributions.}
 to {\bf all} orders of perturbation, either as a bare equation with fixed ultraviolet cutoff,
or as an equation for the renormalized theory. In the latter case $\lambda $ should still be understood 
as the bare constant, but reexpressed as a series in powers of $\lambda_{ren}$.

\section{Ward Identities}

We orient the propagators from a $\bar{\phi}$ to a $\phi$.
For a field $\bar{\phi}_{a b}$ we call the index $a$ a 
{\it left index} and the index, $b$ a {\it right index}. The first (second) index of a $\bar{\phi}$ {\it allways} contracts with 
the second (first) index of a $\phi$.  Consequently for $\phi_{c d}$, $c$ is a {\it right index} and $d$ is a {\it left index}.

Let $U=e^{\imath B}$ with $B$ a small hermitian matrix. We consider the ``left" (as it acts only on the left indices) change of variables\footnote{There is a similar ``right" change of variables, acting only on the right indices.}:
\bea
\phi^U=\phi U;\bar{\phi}^U=U^{\dagger}\bar{\phi} \ .
\eea
The variation of the action is, at first order:
\bea
\delta S&=&\phi U X U^{\dagger}\bar{\phi}-\phi X \bar{\phi}\approx
\imath\big{(}\phi B X\bar{\phi}-\phi X B \bar{\phi}\big{)}\nonumber\\
&=&\imath B\big{(}X\bar{\phi}\phi-\bar{\phi}\phi X \big{)}
\eea
and the variation of the external sources is:
\bea
\delta F&=&U^{\dagger}\bar{\phi}\eta-\bar{\phi}\eta+\bar{\eta}\phi U-\bar{\eta}\phi 
        \approx-\imath B \bar{\phi}\eta+\imath\bar{\eta}\phi B\nonumber\\
	&=&\imath B\big{(}-\bar{\phi}\eta+\bar{\eta}\phi{)} .
\eea
We obviously have:
\bea
&&\frac{\delta \ln Z}{\delta B_{b a}}=0=\frac{1}{Z(\bar{\eta},\eta)}\int d\bar{\phi} d\phi
   \big{(}-\frac{\delta S}{\delta B_{b a}}+\frac{\delta F}{\delta B_{b a}}\big{)}e^{-S+F}\nonumber\\
   &&=\frac{1}{Z(\bar{\eta},\eta)}\int d\bar{\phi} d\phi  ~e^{-S+F}
\big{(}-[X \bar{\phi}\phi-\bar{\phi}\phi X]_{a b}+
       [-\bar{\phi}\eta+\bar{\eta}\phi]_{a b}\big{)} \  .
\eea

We now take $\partial_{\eta}\partial_{\bar{\eta}}|_{\eta=\bar{\eta}=0}$ 
on the above expression. As we have at most two insertions we get only the connected components of the correlation functions.
\bea
0=<\partial_{\eta}\partial_{\bar{\eta}}\big{(}
-[X \bar{\phi}\phi-\bar{\phi}\phi X]_{a b}+
       [-\bar{\phi}\eta+\bar{\eta}\phi]_{a b}\big{)}e^{F(\bar{\eta},\eta)} |_0>_c \ ,
\eea
which gives:
\bea
<\frac{\partial(\bar{\eta}\phi)_{a b}}{\partial \bar{\eta}}\frac{\partial(\bar{\phi}\eta)}{\partial \eta}
-\frac{\partial(\bar{\phi}\eta)_{a b}}{\partial \eta}\frac{\partial (\bar{\eta}\phi)}{\partial \bar{\eta}}
- [X \bar{\phi}\phi-\bar{\phi}\phi X]_{a b}
\frac{\partial(\bar{\eta}\phi)}{\partial \bar{\eta}}\frac{\partial (\bar{\phi}\eta)}{\partial\eta}>_c=0 .
\eea
Using the explicit form of $X$ we get:
\bea
(a-b)<[ \bar{\phi}\phi]_{a b}
\frac{\partial(\bar{\eta}\phi)}{\partial \bar{\eta}}\frac{\partial (\bar{\phi}\eta)}{\partial\eta}>_c=
<\frac{\partial(\bar{\eta}\phi)_{a b}}{\partial \bar{\eta}}\frac{\partial(\bar{\phi}\eta)}{\partial \eta}>_c
-<\frac{\partial(\bar{\phi}\eta)_{a b}}{\partial \eta}\frac{\partial (\bar{\eta}\phi)}{\partial \bar{\eta}}> \ ,
\nonumber
\eea
and for $\bar{\eta}_{ \beta \alpha} \eta_{ \nu \mu}$ we get:
\bea
(a-b)<[ \bar{\phi}\phi]_{a b} \phi_{\alpha \beta} 
\bar{\phi}_{\mu \nu }>_c=
<\delta_{a \beta}\phi_{\alpha b} \bar{\phi}_{\mu \nu}>_c
-<\delta _{b \mu }\bar{\phi}_{a \nu} \phi_{\alpha \beta}>_c
\eea

We now restrict to terms in the above expressions which are planar with a single external face,
as all others are irrelevant. Such terms have $\alpha=\nu$, $a=\beta$ and $b=\mu$. 
The Ward identity for $2$ point function reads:
\bea\label{ward2point}
(a-b)<[ \bar{\phi}\phi]_{a b} \phi_{\nu a} 
\bar{\phi}_{b \nu }>_c=
<\phi_{\nu b} \bar{\phi}_{b \nu}>_c
-<\bar{\phi}_{a \nu} \phi_{\nu a}>_c
\eea
(repeated indices are not summed). 

\begin{figure}[hbt]
\centerline{
\includegraphics[width=100mm]{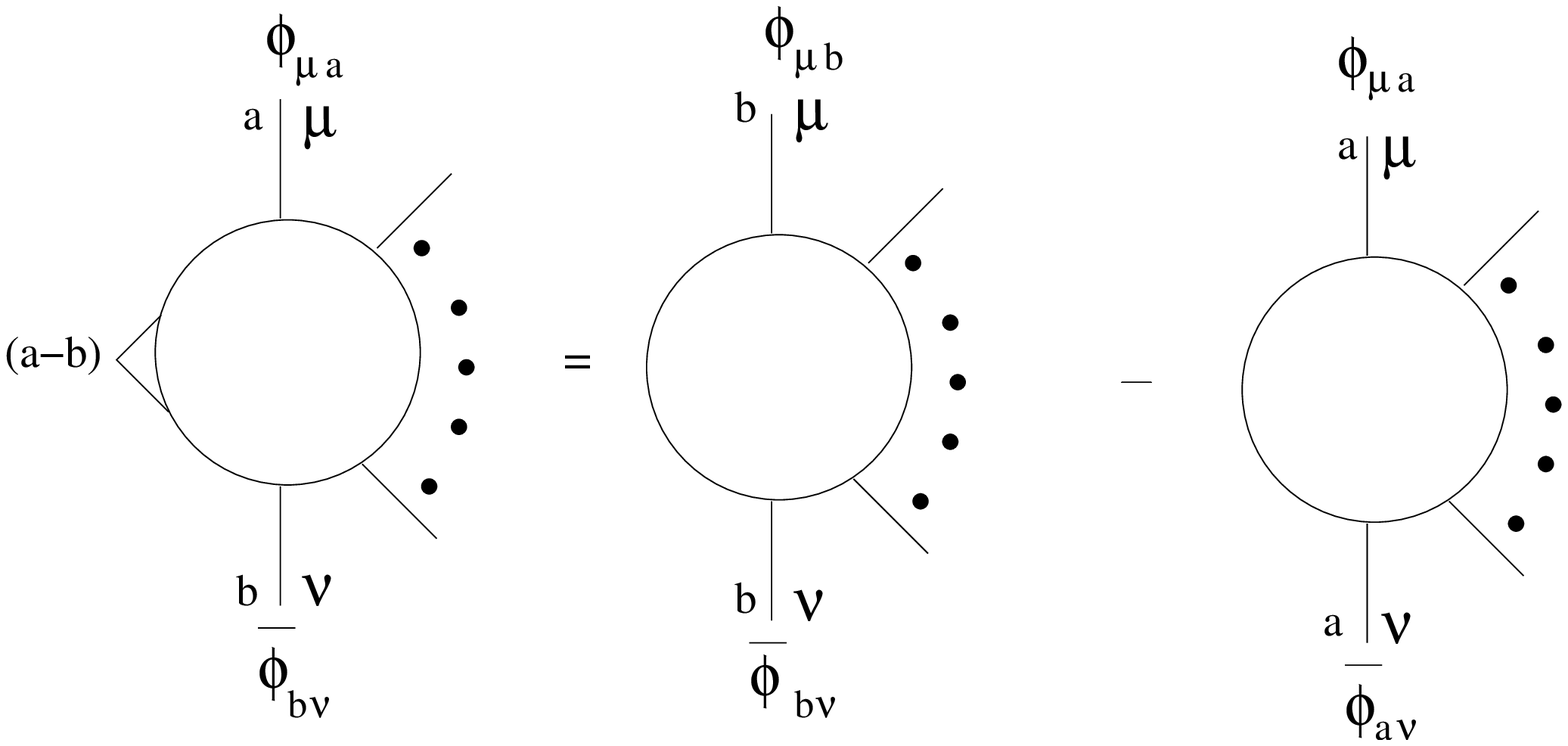}
}
\caption{The Ward identity for a 2p point function with insertion on the left face}\label{fig:Ward}
\end{figure}

The indices $a$ and $b$ are left indices, so that we have the Ward identity with an insertion on a left face\footnote{There is a similar Ward identity obtained with the ``right" transformation, consequently with the insertion on a right face.}  as represented in  Fig. \ref{fig:Ward}.

\section{Proof of the Theorem}

We start this section by some definitions:
we will denote $G^{4}(m,n,k,l)$ the connected four point function restricted to the planar one broken face case, where $m,n,k,l$  are the indices of the external face in the correct cyclic order. The first index $m$ allways represents a left index.

\begin{figure}[hbt]
\centerline{
\includegraphics[width=60mm]{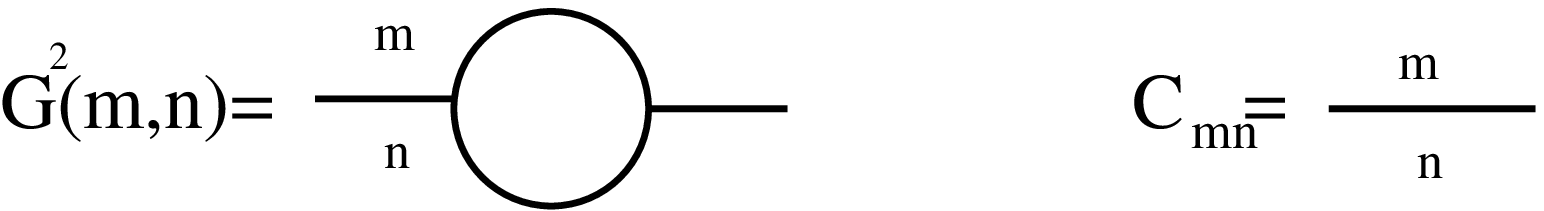}
}
\caption{The {\bf dressed} and the bare propagators}\label{fig:propagators}
\end{figure}

Similarely, $G^{2}(m,n)$ is the connected planar one broken face two point function with $m,n$ the indices on the external face (also called the {\bf dressed} propagator, see Fig. \ref{fig:propagators}). $G^{2}(m,n)$ and $\Sigma(m,n)$ are related by:
\bea
\label{G2Sigmarelation}
  G^{2}(m,n)=\frac{C_{m n}}{1-C_{m n}\Sigma(m,n)}=\frac{1}{C_{m n}^{-1}-\Sigma(m,n)} \, .
\eea 

$G_{ins}(a,b;...)$ will denote the planar one broken face 
connected function with one insertion on the left border where the matrix index jumps from $a$ to $b$. With this notations the Ward identity (\ref{ward2point}) writes:
\bea
(a-b) ~ G^{2}_{ins}(a,b;\nu)=G^{2}(b,\nu)-G^{2}(a,\nu)\, .
\eea

All the identities we use, either Ward identities or the Dyson equation of motion
can be written either for the bare  theory or for the theory with complete mass renormalization, which is the one considered in \cite{DR}. In the first case the parameter $A$ in (\ref{propafixed}) is the bare one, $A_{bare}$
and there is no mass subtraction. In the second case the parameter $A$ in (\ref{propafixed}) 
is $A_{ren}= A_{bare} - \Sigma(0,0)$, and every two point 1PI subgraph is subtracted at 0 external indices\footnote{These mass subtractions need not be rearranged into forests 
since 1PI 2point subgraphs never overlap non trivially.}. Troughout this paper $\partial_{L}$ will denote the derivative with respect to a left index and $\partial_{R}$ the one with respect to a right index. When the two derivatives are equal we will employ the generic notation $\partial$. 

Let us prove first the Theorem in the mass-renormalized case, then in the next subsection
in the bare case. Indeed the mass renormalized theory used is free from any quadratic divergences, and remaining logarithmic subdivergences in the ultra violet  cutoff can be removed easily by going, for instance, to the ``useful" renormalized effective series, 
as explained in \cite{DR}. 

\begin{figure}[hbt]
\centerline{
\includegraphics[width=120mm]{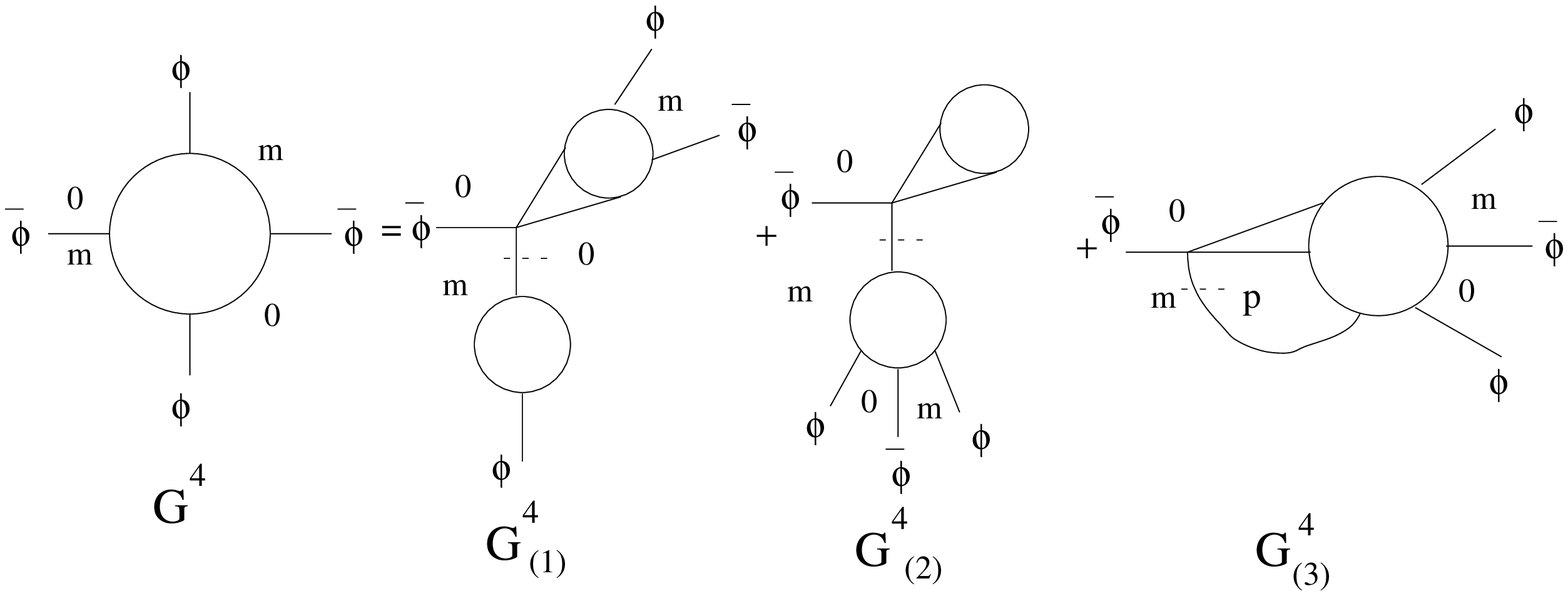}
}
\caption{The Dyson equation}\label{fig:dyson}
\end{figure}

We analyze a four point connected function $G^4(0,m,0,m)$ with index $m \ne 0$ on the right borders. 
This explicit break of left-right symmetry is adapted to our problem.

Consider a $\bar{\phi}$ external line and the first vertex hooked to it. 
Turning right on the $m$ border at this vertex we meet a new line (the slashed line in Fig. \ref{fig:dyson}). The slashed line either separates the graph into two disconnected components ($G^{4}_{(1)}$ and $G^{4}_{(2)}$ in Fig. \ref{fig:dyson}) or not 
($G^{4}_{(3)}$ in Fig. \ref{fig:dyson}). Furthermore, if the slashed line separates the graph into two disconnected components the first vertex may either belong to a four point component ($G^{4}_{(1)}$ in Fig. \ref{fig:dyson}) or to a two point component
($G^{4}_{(2)}$ in Fig. \ref{fig:dyson}). 

We stress that this is a {\it classification} of graphs: the different components depicted in Fig. \ref{fig:dyson} take into account all the combinatoric factors. Furthermore, the setting of the external indices to $0$ on the left borders and $m$ on the right borders distinguishes the $G^{4}_{(1)}$ and $G^{4}_{(2)}$ from their counterparts ``pointing upwards": indeed, the latter are classified in $G^{4}_{(3)}$!

We have thus the Dyson equation:
\bea
\label{Dyson}
 G^4(0,m,0,m)=G^4_{(1)}(0,m,0,m)+G^4_{(2)}(0,m,0,m)+G^4_{(3)}(0,m,0,m)\, .
\eea    

The second term,  $G^{4}_{(2)}$, is zero. Indeed the mass renormalized two point insertion is zero, as it has the external left index set to zero. Note that this is an insertion exclusively on the left border. The simplest case of such an insertion is a
 (left) tadpole. We will (naturally) call a general insertion touching only the left border a ``generalized left tadpole" and denote it by 
 $T^L$. 

We will prove that $G^{4}_{(1)}+G^{4}_{(3)}$ yields 
$\Gamma^4=\lambda (1-\partial \Sigma)^2$ after amputation of the four external propoagators.

We start with $G^{4}_{(1)}$. It is of the form:
\bea
G^4_{(1)}(0,m,0,m)=\lambda C_{0 m} G^{2}(0, m) G^{2}_{ins}(0,0;m)\,.
 \eea

By the Ward identity we have:
\bea
G^{2}_{ins}(0,0;m)&=&\lim_{\zeta\rightarrow 0}G^{2}_{ins}(\zeta ,0;m)=
\lim_{\zeta\rightarrow 0}\frac{G^{2}(0,m)-G^{2}(\zeta,m)}{\zeta}\nonumber\\
&=&-\partial_{L}G^{2}(0,m) \, .
\eea
Using the explicit form of the bare propagator we have $\partial_L C^{-1}_{ab}=\partial_R C^{-1}_{ab}=\partial C^{-1}_{ab}=1$. Reexpressing $G^{2}(0,m)$ by eq.  (\ref{G2Sigmarelation}) we conclude that:
\bea\label{g41}
G^4_{(1)}(0,m,0,m)&=&\lambda
C_{0m}\frac{C_{0m}C^2_{0m}[1-\partial_{L}\Sigma(0,m)]}{[1-C_{0m}\Sigma(0,m)]
(1-C_{0m}\Sigma(0,m))^2}\nonumber\\
&=&\lambda [G^{2}(0,m)]^{4}\frac{C_{0m}}{G^{2}(0,m)}[1-\partial_{L}\Sigma(0,m)]\, .
\eea
The self energy is (again up to irrelevant terms (\cite{GW2}):
\bea
\label{PropDressed}
\Sigma(m,n)=\Sigma(0,0)+(m+n)\partial\Sigma(0,0) 
\eea 
Therefore up to irrelevant terms ($C^{-1}_{0m}=m+A_{ren}$) we have:
\bea
\label{G2(0,m)}
G^{2}(0,m)=\frac{1}{m+A_{bare}-\Sigma(0,m)}=\frac{1}{m[1-\partial\Sigma(0,0)]+A_{ren}}
\, ,
\eea
and
\bea \label{cdressed}
\frac{C_{0m}}{G^{2}(0,m)}=1-\partial\Sigma(0,0)+\frac{A_{ren}}{m+A_{ren}}\partial\Sigma(0,0) \, .
\eea
Inserting eq. (\ref{cdressed}) in eq. (\ref{g41}) holds:
\bea
\label{g41final}
G^4_{(1)}(0,m,0,m)&=&\lambda [G^{2}(0,m)]^{4}\bigl( 
1-\partial\Sigma(0,0)+\frac{A_{ren}}{m+A_{ren}}\partial\Sigma(0,0)
\bigr) \nonumber\\
&&[1-\partial_{L}\Sigma(0,m)]\, .
\eea

\begin{figure}[hbt]
\centerline{
\includegraphics[width=140mm]{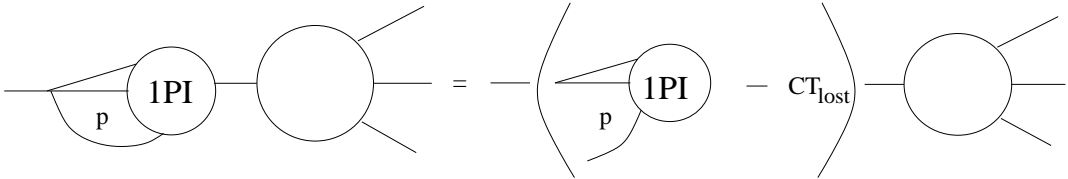}
}
\caption{Two point insertion and opening of the loop with index $p$}\label{fig:insertion}
\end{figure}

For the $G^4_{(3)}(0,m,0,m)$ one starts by ``opening" the face which is ``first on the right''. The summed index of this face is called  $p$ (see Fig. \ref{fig:dyson}).  For bare Green functions this reads:
\bea
\label{opening}
G^{4,bare}_{(3)}(0,m,0,m)=C_{0m}\sum_{ p} G^{4,bare}_{ins}(p,0;m,0,m)\, .
\eea
When passing to mass renormalized Green functions one must be cautious. It is possible that the face $p$ belonged to a  1PI two point insertion in $G^{4}_{(3)}$ (see the left hand side in Fig. \ref{fig:insertion}).
Upon opening the face $p$ this 2 point insertion disappears (see right hand side of Fig. \ref{fig:insertion})! 
When renormalizing, the counterterm  corresponding to this kind of two point insertion will be substracted on the left hand side of  eq.(\ref{opening}), but not on the right hand side. In the equation for $G^{4}_{(3)}(0,m,0,m)$ one must 
therefore \textit{add its missing counterterm}, so that:
\bea
\label{Open2}
G^4_{(3)}(0,m,0,m)&=& C_{0m}\sum_{p} G^{4}_{ins}(0,p;m,0,m)\nonumber\\
     &-&C_{0m}(CT_{lost})G^{4}(0,m,0,m)\,.
\eea

\begin{figure}[hbt]
\centerline{
\includegraphics[width=90mm]{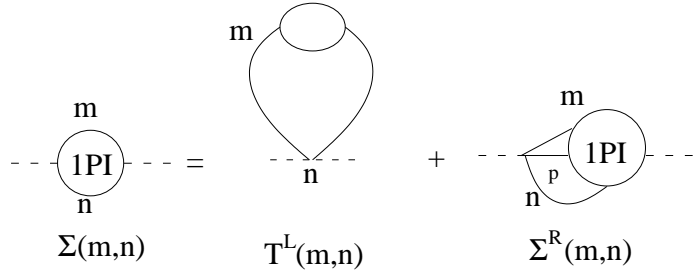}
}
\caption{The self energy}\label{fig:selfenergy}
\end{figure}

It is clear that not all 1PI 2 point insertions on the left hand side of Fig. \ref{fig:insertion} will be ``lost" on the right hand side. If the insertion is a ``generalized left tadpole" it is not ``lost" by opening the face $p$ (imagine a tadpole pointing upwards in Fig.\ref{fig:insertion}: clearely it will not be opened by opening the line). We will call the 2 point 1PI insertions ``lost" on the right hand side $\Sigma^R(m,n)$. Denoting the generalized left tadpole $T^{L}$ we can write (see Fig .\ref{fig:selfenergy}):
\bea
\label{eq:leftright}
  \Sigma(m,n)=T^{L}(m,n)+\Sigma^R(m,n)\, .
\eea
Note that as $T^{L}(m,n)$ is an insertion exclusively on the left border, it does not depend upon the right index $n$. We therefore have $\partial\Sigma(m,n)=\partial_R\Sigma(m,n)=\partial_R\Sigma^R(m,n)$.

The missing mass counterterm writes:
\bea\label{lostct}
CT_{lost}=\Sigma^R(0,0)=\Sigma(0,0)-T^{L}\, .
\eea
In order to evaluate $\Sigma^{R}(0,0)$ we proceed
by opening its face $p$ and using the Ward identity (\ref{ward2point}), to obtain:
\bea
\label{S2}
\Sigma^R(0,0)&=&\frac{1}{G^{2}(0,0)}\sum_{p}G^2_{ins}(0,p;0)\nonumber\\
    &=&\frac{1}{G^{2}(0,0)}\sum_{p}\frac{1}{p}[G^2(0,0)-G^2(p,0)]\nonumber\\
    &=&\sum_{p}\frac{1}{p} \biggl(1 -\frac{G^{2}(p,0)}{G^{2}(0,0)}\biggr) \, .
\eea

Using eq. (\ref{Open2}) and eq. (\ref{S2}) we have:
\bea\label{S3}
G^4_{(3)}(0,m,0,m)&=& C_{0m}\sum_{p} G^{4}_{ins}(0,p;m,0,m)\nonumber\\
&-&C_{0m} G^{4}(0,m,0,m) \sum_{p}\frac{1}{p}  
\biggl( 1- \frac{G^{2}(p,0)}{G^2(0,0)}  \biggr)  \, .
\eea

After some manipulations using mainly the Ward identity, detiled in \cite{beta} we obtain the final result
\bea\label{g43}
G^4_{(3)}(0,m,0,m)&=&-C_{0m}G^{4}(0,m,0,m)\frac{1}{G^{2}(0,0)}\frac{\partial\Sigma(0,0)}{1-\partial\Sigma(0,0)}
\nonumber\\
&=&-G^{4}(0,m,0,m)
\frac{A_{ren} \; \partial\Sigma(0,0)}{(m+A_{ren}) [1-\partial\Sigma(0,0)]} \ .
\eea
Using (\ref{g41final}) and (\ref{g43}), equation (\ref{Dyson}) rewrites as:
\bea
\label{final}
&&G^4(0,m,0,m)\Big{(}1+
\frac{A_{ren}\; \partial\Sigma(0,0)}{(m+A_{ren}) \; [ 1-\partial\Sigma(0,0)] }\Big{)}
\\
&&=\lambda (G^{2}(0,m))^{4}\Big{(}1-\partial\Sigma(0,0)+\frac{A_{ren}}{m+A_{ren}}\partial\Sigma(0,0)\Big{)}
[1-\partial_{L}\Sigma(0,m)]\, .\nonumber
\eea
We multiply (\ref{final}) by $[1-\partial\Sigma(0,0)]$ and amputate four times. As the differences $\Gamma^4(0,m,0,m,)-\Gamma^4(0,0,0,0)$ and $\partial_L\Sigma(0,m)-\partial_L\Sigma(0,0)$ are irrelevant we get:
\bea
\Gamma^{4}(0,0,0,0)=\lambda (1-\partial\Sigma(0,0))^2\, .
\eea
\qed 

\section{Conclusion}

Since the main result of this paper is proved up to irrelevant 
terms which converge at least like a power of the ultraviolet cutoff, as this ultraviolet cutoff is lifted towards infinity,
we not only get that the beta function vanishes in the ultraviolet regime, but that it 
vanishes fast enough so that the total flow of the coupling constant is bounded. 
The reader might worry whether this conclusion is still true for the full model which has 
$\Omega_{ren} \ne 1$, hence no exact conservation of matrix indices along faces.
The answer is yes, because the flow of $\Omega $ towards its ultra-violet limit
$\Omega_{bare}=1$ is very fast (see e.g. \cite{DR}, Sect II.2). 

The vanishing of the beta function is a step towards a full non perturbative construction 
of this model without any cutoff, just like e.g. the one of the Luttinger model \cite{BGPS,BM1}.
But NC $\phi^4_4$ would be the first such \textit{four dimensional} model, and the only one
with non logarithmic divergences. Tantalizingly, quantum field theory might actually behave 
better and more interestingly on non-commutative than on commutative spaces.
Steps in this directions have been taken in \cite{const1,const2}.


\medskip

\begin{thebibliography}{99}

\bibitem{beta}
  M.~Disertori, R.~Gurau, J.~Magnen and V.~Rivasseau,
  ``Vanishing of beta function of non commutative phi(4)**4 theory to all
  orders,''
  Phys.\ Lett.\  B {\bf 649}, 95 (2007) [arXiv:hep-th/0612251].

\bibitem{DN}
M. Douglas and  N. Nekrasov, ``Noncommutative field theory,''
Reviews of Modern Physics, {\bf 73}, 9771029 (2001)

\bibitem{CDS} A.~Connes, MR.~Douglas, A.~Schwarz
``Noncommutative Geometry and Matrix Theory: Compactification on Tori", 
JHEP {\bf 9802} (1998) 003 [arXiv:hep-th/9711162].

\bibitem{SW}
N.~Seiberg and E.~Witten, ``String theory and noncommutative geometry,''
JHEP {\bf 9909} (1999) 032 [arXiv:hep-th/9908142].

\bibitem{Connes}
  A.~H.~Chamseddine, A.~Connes and M.~Marcolli,
  ``Gravity and the standard model with neutrino mixing,''
  arXiv:hep-th/0610241.

\bibitem{GW1}
H.~Grosse and R.~Wulkenhaar,
``Power-counting theorem for non-local matrix models and renormalization,''
 Commun.\ Math.\ Phys. {\bf 254}, (2005) 91-127, [arXiv:hep-th/0305066]

\bibitem{GW2} H.~Grosse and R.~Wulkenhaar, ``Renormalization
of $\phi^4$-theory on noncommutative ${\mathbb R}^4$ in the matrix
base,'' Commun.\ Math.\ Phys. {\bf 256}, (2005) 305-374, [arXiv:hep-th/0401128] 

\bibitem{RVW}
V.~Rivasseau, F.~Vignes-Tourneret and R.~Wulkenhaar,
``Renormalization of noncommutative $\phi^{\star4}_4$-theory by multi-scale 
analysis,''  Commun. Math. Phys. {\bf 262}, 565 (2006), [arXiv:hep-th/0501036]

\bibitem{GMRV}
R. Gurau, J. Magnen, V. Rivasseau and F. Vignes-Tourneret
``Renormalization of Non Commutative $\Phi^4_4$ Field Theory in Direct Space,''
Commun.\ Math.\ Phys. {\bf 267}, 515-542 (2006) [arXiv:hep-th/0512271]

\bibitem{LS}
E.~Langmann and R.~J.~Szabo,
``Duality in scalar field theory on noncommutative phase spaces,''
Phys.\ Lett.\ B {\bf 533} (2002) 168
[arXiv:hep-th/0202039].

\bibitem{GR}
R.~Gurau and V.~Rivasseau, Parametric Representation of Noncommutative Field Theory, 
to appear in Commun. Math. Phys., [arXiv:math-ph/0606030] 

\bibitem{dimreg}
  R.~Gurau and A.~Tanasa,
  ``Dimensional regularization and renormalization of non-commutative QFT,''
  arXiv:0706.1147 [math-ph].
  %%CITATION = ARXIV:0706.1147;%%

\bibitem{melin}
  R.~Gurau, A.~P.~C.~Malbouisson, V.~Rivasseau and A.~Tanasa,
  ``Non-Commutative Complete Mellin Representation for Feynman Amplitudes,''
  Lett.\ Math.\ Phys.\  {\bf 81}, 161 (2007)
  [arXiv:0705.3437 [math-ph]].
  %%CITATION = LMPHD,81,161;%%

\bibitem{ConKrTV}
  A.~Tanasa and F.~Vignes-Tourneret,
  ``Hopf algebra of non-commutative field theory,''
  arXiv:0707.4143 [math-ph].
  %%CITATION = ARXIV:0707.4143;%%

\bibitem{propaga}
  R.~Gurau, V.~Rivasseau and F.~Vignes-Tourneret,
  ``Propagators for noncommutative field theories,''
  Annales Henri Poincare {\bf 7}, 1601 (2006)
  [arXiv:hep-th/0512071].
  %%CITATION = AHPJF,7,1601;%%

\bibitem{param2}
  V.~Rivasseau and A.~Tanasa,
  ``Parametric representation of 'critical' noncommutative QFT models,''
  arXiv:math-ph/0701034.
  %%CITATION = MATH-PH/0701034;%%

\bibitem{thooft} G. 't Hooft, ``The Glorious Days of Physics - Renormalization of Gauge theories,'' hep-th/9812203

\bibitem{GWbeta} H.~Grosse and R.~Wulkenhaar,  ``The $\beta$-function in duality-covariant 
non-commutative $\phi^4$-theory,'' Eur. Phys. J. C {\bf 35}, 277-282 (2004), [arXiv:hep-th/0402093]

\bibitem{DR} M. Disertori and V. Rivasseau, ``Two and Three Loops Beta Function of Non Commutative $\Phi^4_4$ Theory,'' hep-th/0610224, to appear in European Physical Journal C.

\bibitem{BM1} G. Benfatto and V. Mastropietro, ``Ward Identities and Chiral Anomaly in the Luttinger Liquid,'' 
Commun. Math. Phys. Vol. {\bf 258}, 609-655 (2005).

\bibitem{BGPS} G. Benfatto, G. Gallavotti, A. Procacci and B. Scoppola,
``Beta function and Schwinger functions for a many fermions system in one dimension. 
Anomaly of the Fermi surface ,"  Commun. Math. Phys. Vol {\bf 160},  93-171, 1994

\bibitem{const1}
  V.~Rivasseau,
  ``Constructive Matrix Theory,''
  arXiv:0706.1224 [hep-th].
  %%CITATION = ARXIV:0706.1224;%%

\bibitem{const2}
  J.~Magnen and V.~Rivasseau,
  ``Constructive $\phi^4$ field theory without tears,''
  arXiv:0706.2457 [math-ph].
  %%CITATION = ARXIV:0706.2457;%%

\end{thebibliography}
\end{document}